\def\bq{\begin{equation}}
\def\eq{\end{equation}}
\def\bqa{\begin{eqnarray}}
\def\eqa{\end{eqnarray}}
\def\bqb{\begin{eqnarray*}}
\def\eqb{\end{eqnarray*}}
\documentstyle[12pt]{article}\textwidth 16cm
\def\pr#1#2#3{ Phys. Rev. ${\bf{#1}}$ (#2) #3}
\def\prl#1#2#3{ Phys. Rev. Lett. ${\bf{#1}}$ (#2) #3}
\def\pl#1#2#3{ Phys. Lett. ${\bf{#1}}$ (#2) #3 }
\def\prep#1#2#3{ Phys. Rep. ${\bf{#1}}$ (#2) #3}
\def\np#1#2#3{ Nucl. Phys. ${\bf{#1}}$ (#2) #3}
\def\zp#1#2#3{ Z. Phys. ${\bf{#1}}$ (#2) #3}

\def\etal{{\it et.al.\/}}

\global\nulldelimiterspace = 0pt

\def\Bsl{\hbox{/\kern-.6700em$B$}} 
\def\Dsl{\hbox{/\kern-.6700em$D$}} 
\def\Wsl{\hbox{/\kern-.6700em$W$}} 

\def\roughly#1{\mathrel{\raise.3ex
    \hbox{$#1$\kern-.75em\lower1ex\hbox{$\sim$}}}}
\def\lsim{\roughly<}
\def\gsim{\roughly>}

\def\ol#1{\overline{#1}}

\def\L{ {\cal L }}
\def\O{ {\cal O }}

\def\swd{s^2_W}

\def\mwd{M_W^2}
\def\mw{M_W}

\def\mh2{m^2_H}
\def\mzd{M^2_Z}

\begin{document}
\pagenumbering{arabic}
\thispagestyle{empty}

\begin{flushright}
 PM/95-20 \\ THES-TP 95/07 \\
May 1995 \end{flushright}
\vspace{2cm}

\begin{center}
{\Large\bf Anomalous Couplings in Single Higgs Production
through $\gamma\gamma$ Collisions}
\footnote{Partially supported by the EC contract CHRX-CT94-0579.}
 \vspace{1.5cm}  \\
{\large G.J. Gounaris$^{a}$ and F.M. Renard$^b$}
\vspace {0.5cm}  \\
$^a$Department of Theoretical Physics, University of Thessaloniki,\\
Gr-54006, Thessaloniki, Greece,\\
\vspace{0.2cm}
$^b$Physique
Math\'{e}matique et Th\'{e}orique,
CNRS-URA 768,\\
Universit\'{e} de Montpellier II,
 F-34095 Montpellier Cedex 5.\\

\vspace{2cm}

 {\bf Abstract}
\end{center}
\noindent
We show that single Higgs production in $\gamma\gamma$
collisions through laser backscattering provides the best way to
look for New Physics (NP) effects inducing anomalous Higgs couplings.
Our analysis is
based on the four $dim=6$ operators $\O_{UB}$, $\O_{UW}$,
$\ol{\O}_{UB}$ and $\ol{\O}_{UW}$ which describe  New
Physics effects in this sector. Using the Higgs
production rate we establish
observability limits for the couplings of the aforementioned
operators at a level of $10^{-3}$,
which means establishing lower bounds on the scale of NP of the
order 30 to 200 TeV. Higgs branching ratios, especially the ratio
$\Gamma(H\to\gamma\gamma)/\Gamma (H\to\gamma Z)$, are shown to provide
powerful ways to disentangle the effects of the various operators.

\vspace{1cm}

\setcounter{page}{0}
\def\thefootnote{\arabic{footnote}}
\setcounter{footnote}{0}
\clearpage

\section{Introduction}

Future high energy linear $e^+e^-$ colliders should allow the
realization of $\gamma\gamma$ collisions with intense high energy
photon beams through the laser backscattering
method \cite{laser}. Provided the Higgs particle will be
accessible at the future colliders, the $\gamma\gamma$ fusion
into a single Higgs boson
has been recognized as being a powerful tool to study the
anomalous Higgs couplings and thereby the scalar
sector of the electroweak interactions \cite{higpro}.
These anomalous Higgs couplings may arise as
residual effects of a New
Physics (NP) characterized by a scale $\Lambda \gg \mw$.\par

It has been shown that there exist seven $dim=6$ $SU(2)\times
U(1)$ gauge
invariant operators which are not strongly constrained by
existing LEP1 experiments and provide a reasonably complete
{}~description of the residual purely bosonic NP interactions
at energies lower
than $\Lambda$ \cite {DeR, blind, model, GRVZbb}.
Furthermore, only four of these operators
dubbed $\O_{UB}$, $\O_{UW}$, $\ol{\O}_{UB}$ and $\ol{\O}_{UW}$,
create Higgs anomalous couplings exclusively and are therefore
not affected by constraints on the anomalous gauge couplings.
A first study of the operator $\O_{UW}$ has already been
done in \cite {ggVV} by looking into vector boson production through
 $\gamma\gamma$  collisions, and subsequently in \cite{higpro}
where $\gamma\gamma\to H$ and the various Higgs branching ratios
have been analysed.\par

In the companion paper \cite{ggVV1} we have  presented
a complete analysis of
what can be learned for all seven operators by using weak boson
pair production at $\gamma \gamma $ colliders.
The aim of the present paper is to perform an analysis of the
process $\gamma \gamma \to H$ and of the Higgs decay modes. These
later processes are particularly sensitive to $\O_{UB}$,
$\O_{UW}$, $\ol{\O}_{UB}$, $\ol{\O}_{UW}$ and can be used to
strongly constrain their couplings.\par

The Standard contribution to $\gamma\gamma\to H$ only occurs at
1-loop.  With the high luminosities expected
at linear $e^+e^-$ colliders, a large number of Higgs bosons
should be produced. The
sensitivity to anomalous $H\gamma\gamma$
couplings is therefore very strong. It turns out that
$\gamma\gamma\to H$ can be used to put an observability limit of
the order of $10^{-3}$ on these couplings, which means that NP
scales of up to several tens of
TeV can be probed. This range largely covers the domain of characteristic
scales expected by several theoretical models.\par

The disentangling of the four operators $\O_{UB}$,
$\O_{UW}$, $\ol{\O}_{UB}$ and $\ol{\O}_{UW}$
is further helped by considering the Higgs decay branching
ratios. Concerning this we remark that $H\to b\bar
b$ is not affected by the aforementioned seven purely bosonic
operators descibing NP. Neither $H\to
WW,\, ZZ$ are particularly sensitive to such an NP, since it is masked by
strong tree level SM contributions. On the other hand, the processes
$H \to \gamma\gamma$ and  $H \to \gamma Z$ who receive tree level
contributions from the above anomalous couplings and only one
loop ones from SM, are most sensitive to NP.
Thus their ratio to the dominant Higgs decay
mode, which depending on the Higgs mass may either be the $b\bar b$
or the $WW,\, ZZ$ modes, provide a very sensitive way to further help
disentangling among the CP conserving operators $\O_{UB}$ and
$\O_{UW}$. This way, using $H \to \gamma \gamma$ or $H \to
\gamma Z$,  couplings even weaker than $10^{-3}$ could
be observable. For comparison we note that the corresponding
sensitivity limit from $H\to WW,\, ZZ$ is at the 10\% level.
The identification of the CP-violation effects
and the disentangling of the CP-violating operators from
the CP-conserving ones requires a study of either the W,Z spin
density matrix through their decay distributions or of initial
$\gamma\gamma$ linear polarization effects.  \par

In Section 2 we give the explicit expression of the
production rate of Higgs through $\gamma\gamma$ collisions at a
linear collider, including SM and NP contributions due to the
four operators. We consider collider energies of 0.5, 1 and 2
TeV. Section 3 is devoted to the study of the Higgs  decay
modes $H\to \gamma\gamma$, $\gamma Z$, $WW$, $ZZ$, $b\bar b$
and to the computation of the SM and NP effects for the various
ratios. A discussion of the sensitivities to the anomalous
couplings and of the possibility to disentangle the
contributions from the various operators is given in Section 4.\par

\section{The $\gamma\gamma\to H$ production rate}

The formalism of $\gamma\gamma$ collisions through laser
backscattering has been described in \cite{laser}. We shall use
the same notations as in \cite{higpro}. The cross section for
$\gamma\gamma\to H$ is given by

\bq
\sigma  = \L_{\gamma \gamma}(\tau_H)~\left({8\pi^2 \over
 m_H} \right)~{\Gamma(H\to\gamma \gamma)\over s_{ee}}
    \ \ \ \ \ \ \ , \ \ \ \ \ \
\eq
where the luminosity function $\L_{\gamma\gamma}(\tau_H)$ for
$\tau_H = {m^2_H / s_{ee}}$ is ~explicitly given in terms
of the $f^{laser}_{\gamma/e}$ distribution in \cite{higpro, ggVV}.
Essentially $\L_{\gamma\gamma}$ is close to the $e^+e^-$ linear collider
luminosity $\L_{ee}$ up to $\tau_{max}=(0.82)^2$.\par

The $H\to \gamma\gamma$ decay width is computed in terms of the
one-loop SM contribution and tree level NP ones \cite{review}:
\bq
 \Gamma(H\to\gamma \gamma) = {\sqrt2 G_F \over 16\pi} m^3_H
\left (\Big |{\alpha\over4\pi}({4\over3}F_t + F_W) - 2ds^2_W-2d_B c^2_W
\Big |^2  +4|\bar ds^2_W+\bar d_B c^2_W|^2 \right )\ , \ \ \ \ \
\eq
where the standard top and $W$ contributions are respectively determined
by
\bq
F_t = -2t_t(1+(1-t_t)f(t_t))   \ \ \ \ \ , \ \ \ \ \
\eq
\bq
F_W = 2+3t_W+3t_W(2-t_W)f(t_W)) \ \ \ \ \ \ , \ \ \ \
\eq
in terms of
\bqa
f(t) = \left [sin^{-1}(1/\sqrt{t})\right]^2 \ \ \ \ \ \ \ \
\ \ &
\makebox{\ \ \ \ \ if \ \ \ } &
           t \geq 1 \ \ \ \ , \ \ \ \nonumber \\[0.5cm]
f(t)= -{1\over4}\left [\ln \left ({1+\sqrt{1-t} \over 1-\sqrt{1-t}}
\right )
-i\pi\right ]^2 & \makebox{\ \ \ \ if \ \ \ } & t < 1 \ \
\ \ , \ \ \ \
\eqa
with $t_t =4m^2_t/\mh2$\, , \, $t_W=4\mwd/\mh2$\, and
$m_t=175 GeV$ \cite{CDFD0}.
The NP contributions are obtained from the Lagrangian given in
eqs. (8) in the companion paper, where also the definitions of the
various operators are given.\par

The resulting rate is shown in Fig.1-2 for three typical NLC
energies, 0.5, 1 and 2 TeV. The number of events indicated
in these figures corresponds to $e^+e^-$
luminosities of 20, 80 and 320 $fb^{-1}$
respectively. Strong interference effects may appear (depending
on the Higgs mass) between the SM and the CP-conserving NP
contributions; (compare Fig.1). But
in the case of CP-violating contributions (Fig.2) there are
never such ~interferences. In the figures we have
only illustrated the cases of $\O_{UW}$ and $\ol{\O}_{UW}$. The
corresponding results for $\O_{UB}$ and $\ol{\O}_{UB}$ can be
deduced according to eq(2) from Fig.1-2, by the replacement $d\,
[\bar d] \to {c^2_W/ s^2_W}d_B\, [\bar d_B]$.
So the sensitivity to these last two
operators is enhanced by more than a factor 3 in the cross
section.\par

With the aforementioned designed luminosities, one gets a few thousands
of Higgs bosons produced in the light or intermediate mass range.
Assuming conservatively an
experimental detection accuracy of about 10\% on the
production rate, one still
gets an observability limit of the order of $10^{-3}$,
$4.10^{-3}$, $3.10^{-4}$, $10^{-3}$
for $d$, $\bar d$,
$d_B$ and $\bar d_B$ respectively. The corresponding constraints
on the NP scale derived on the basis of the unitarity relations
\cite{unit, uni2, model},
are 200, 60, 60 and 30 TeV respectively.\par

\section{Higgs decay widths and ratios}

The expression of the $H \to \gamma\gamma$ width has
been written above in (2). Correspondingly, the $H\to\gamma Z$
width is also expressed in terms of a
1-loop SM contribution and of the NP contributions from the four
operators \cite{review}:
\bqa
\Gamma(H\to \gamma Z) & = & {\sqrt2G_Fm^3_H\over 8 \pi}\Big (1-{\mzd
\over
\mh2}\Big )^3\Bigg (\Big
|{\alpha\over 4\pi}(A_t+A_W)+2(d-d_B)s_Wc_W\Big |^2
\nonumber \\
& \null & \ \  + \, 4s^2_W
c^2_W|\bar d-\bar d_B|^2 \Bigg)  \ , \ \
\eqa
\bq
A_t =
{(-6+16s^2_W)\over 3s_Wc_W}[I_1(t_f,l_t)-I_2(t_t,l_t)]
\ \ \ \  , \ \ \ \ \ \
\eq
\bq
A_W
=-cot\theta_W[4(3-tan^2\theta_W)I_2(t_W,l_W)+[(1+{2\over
t_W})
tan^2\theta_W-(5+{2\over t_W})]I_1(t_W,l_W)] \ ,
\eq
where $t_t =4m^2_t/\mh2$, $t_W=4\mwd/\mh2$ as before, and
$l_t=4m^2_t/ \mzd$ , $l_W=4\mwd/ \mzd$. In (7, 8) the
definitions\footnote{There is a discrepancy in the relative sign
of the $g(x)$ and $f(x)$ terms between the first paper in
\cite{review} and the remaining two. Here we follow the results
of the later two papers.}
\bq
I_1(a,b)={ab\over 2(a-b)}+{a^2b^2\over
2(a-b)^2}[f(a)-f(b)]+{a^2b\over(a-b)^2}[g(a)-g(b)] \  ,
\eq
\bq
   I_2(a,b)=-{ab\over2(a-b)}[f(a)-f(b)]\ \ \ \ , \
\eq
where $f(t)$ is given in (5) and
\bqa
\ \ \ \ \ g(t)=\sqrt{t-1}sin^{-1}({1\over\sqrt{t}}) &
\makebox{ \ \ \ if \ \ \ }  &
t\geq 1 \ \ \ , \ \nonumber \\[.2cm]
g(t) ={1\over2}\sqrt{1-t}\left[ln\left({1+\sqrt{1-t}
\over1-\sqrt{1-t}}\right)-i\pi\right] & \makebox{\ \ \ if \ \ \ }
& t< 1 \ \ \ \ \ . \ \ \
\eqa \par

The $H\to W^+W^-, ZZ$ widths receive a tree level SM contribution
which interfere with the ones from the CP-conserving operators $\O_{UB},
\O_{UW}$. In addition there exist  quadratic contributions
from the CP-violating operators $\ol{\O}_{UB}$, $\ol{\O}_{UW}$.
 For $m_H>2M_V$, $V=W, Z$ one gets
\bqa
\Gamma(H\to VV) & = &
 C_V \left( {\alpha  \beta_V \over 4 \swd m_H}\right )
 \Bigg [3M^2_V-m^2_H+ {m^4_H\over 4M^2_V}
 + 4(d_{VV})^2 \Big \{3M^2_V-2m^2_H+{m^4_H\over 2M^2_V} \Big \}
\nonumber \\
& \null &
 -\ 6d_{VV}(\mh2-2M^2_V) +2(\bar d_{VV})^2 (m^2_H-4M^2_Z)
{m^2_H\over M^2_V} \Bigg ]
\ ,
\eqa
with $C_W=1$, $C_Z=1/(2c^2_W)$, $d_{WW}=d$ and $d_{ZZ}=dc^2_W+d_B
s^2_W$.\par

For $M_V<m_H<2M_V$, we have computed the Higgs decay width with
one gauge boson being virtual and decaying into lepton and
quark pairs. The expressions are
\bq
\Gamma (H \to W^*W)~=~\frac{3\alpha^2m_H}{32 \pi s^4_W}~
[D_{SM}(x)+dD_1(x)
+8d^2D_2(x)+8\bar d^2D_3]
\ \ \ \ \ ,\ \ \ \ \
\eq
\bqa
\Gamma (H \to Z^*Z)& = & \frac{\alpha^2m_H}{128 \pi s^4_W c^4_W}~
\left(7-\frac{40\swd}{3}+\frac{160s^4_W}{9}\right)
[D_{SM}(x)+(d_{ZZ})D_1(x)  \nonumber \\
  & \null & + \ 8(d_{ZZ})^2D_2(x)+8(\bar d_{ZZ})^2D_3(x)]
\ \ ,
\eqa
where
\bqa
D_{SM}(x) &= &\frac{3(20x^2-8x+1)}{\sqrt{4x-1}}cos^{-1}
\left(\frac{3x-1}{2x^{3/2}}\right) \nonumber \\
& \null &
  - ~(1-x)\left(\frac{47x}{2} -
\frac{13}{2} +\frac{1}{x} \right) -
3(2x^2-3x+\frac{1}{2})lnx \ \ \ \ \
\ , \ \eqa\\
\bqa
D_1(x) &=&\frac{24(14x^2-8x+1)}{\sqrt{4x-1}}cos^{-1}
\left(\frac{3x-1}{2x^{3/2}}\right) \nonumber \\
 & \null & + ~12(x-1)(9x-5)-12(2x^2-6x+1)lnx \ \ \ \ \ , \ \ \
\eqa\\
\bqa
D_2(x) &= &\frac{54x^3-40x^2+11x-1}{x\sqrt{4x-1}}cos^{-1}
\left(\frac{3x-1}{2x^{3/2}}\right) \nonumber \\
& \null &
+~ \frac{(x-1)}{6}(89x-82+\frac{17}{x})-(3x^2-15x+\frac{9}{2}-
\frac{1}{2x})lnx
\ \ \ \ \ ,
\eqa
\bqa
D_3(x) &= &\frac{-28x^2+11x-1)}{x\sqrt{4x-1}}cos^{-1}
\left(\frac{3x-1}{2x^{3/2}}\right) \nonumber \\
& \null &
  - ~\frac{x^2}{6} -
\frac{21x}{2} +\frac{27}{2} -\frac{17}{6x} +
{(6x^2-9x+1)\over 2x}lnx \ \ \ \ \
\ , \
\eqa
and $x=(M_V/m_H)^2$ with $M_V=M_W$ or $M_Z$.
These results are useful
for $90GeV\lsim m_H \lsim 140 GeV$.
Finally we quote the $H\to b\bar b$ decay
width, which is purely standard and particularly important if
$m_H<140 GeV$. It is given by
 \bq
\Gamma(H\to b\bar b)= 3{\sqrt2G_F m^2_b\over8\pi}\beta^3_b m_H \ \
 \ \ \ , \ \ \
\eq
with $\beta_b = \sqrt{1-{4m^2_b/m^2_H}}$.\par

In Figs. 3a,b the ratios $\Gamma(H\to \gamma\gamma)/\Gamma(H\to b\bar b)$ and
$\Gamma(H\to \gamma Z)/\Gamma(H\to b\bar b)$ are plotted versus $m_H$
for a given value of $d$ or $\bar d$. The case of $d_B$
or $\bar d_B$ can be obtained by the respective replacements
 $d \to d_B c^2_W / s^2_W$ for the $\gamma\gamma$ amplitude
and of $d \to d_B$ for the $\gamma Z$ one. The $m_H$ and $d$
dependences  of
the $\gamma\gamma/b\bar b$ ratio are obviously similar
to the ones of the production cross section
$\sigma(\gamma\gamma\to H)$. The ratios $\gamma Z/b\bar b$
and $\gamma\gamma/\gamma Z$ have independent
 features that may help disentangling the various couplings.
They can also be seen from the ratio shown in Fig.3c.\par

The sensitivity of the $WW/b\bar b$ and $ZZ/b\bar b$ ratios
is much weaker as expected from the ~occurrence of tree level
SM contributions. Because of this these ratios are only useful for
$d \gsim 0.1$. The ratios $\gamma\gamma/WW$ and $\gamma\gamma/ZZ$
are shown in Figs.3d,e and present
the same features as the ratio $\gamma \gamma/ b\bar b$. They can however
be useful in the range of $m_H$ where the $WW, ZZ$ modes are
dominant.\par

Finally we discuss the ratios $WW/ZZ$ and $\gamma\gamma/\gamma
Z$ for a given value of $m_H$ (chosen as $0.2 TeV$ in
the illustrations made in Fig.4a,b),
versus the coupling constant values of
the four operators. This shows very ~explicitly how these
ratios can be used for disentangling the various operators.
They have to be taken in a complementary way to the
$\sigma(\gamma\gamma\to H)$ measurement.
As in the $WW/b \bar b$ case, the $WW/ZZ$ ratio is only useful
for $d \gsim 0.1$, while $\gamma\gamma/\gamma Z$
ratio allows for disentangling $d$ values down to $10^{-3}$
or even less. The corresponding sensitivity limit for $d_B,\,
\bar d_B$ should then be lying at the level of a few times
$10^{-4}$. \par

\section{Final discussion}

At a linear $e^+e^-$ collider the process $\gamma\gamma\to H$
is a very efficient way to produce and study the Higgs boson.
We showed that the sensitivity limits on the NP couplings are at
the $10^{-3}$ level. Using unitarity relations in \cite{unit, uni2,
model} we find and that this implies that new physics scales
are in the range of 30 to 200 TeV, depending on the nature of the
NP operator. \par

In more detail, we have studied the behaviour of
$\sigma(\gamma\gamma\to H)$
versus $m_H$ for the four NP operators which are the candidates
to describe residual NP effects in the Higgs sector.
Ways to disentangle the effects of these operators have been
considered. We have found that this can be achieved by looking at the Higgs
branching ratios into $WW$, $ZZ$, $\gamma\gamma$, $\gamma Z$, which
react rather differently to the presence of each of these operators.
This is illustrated with the ratios of these channels to the $b\bar b$
one which is unaffected by this kind of NP.
Most spectacular for the disentangling of the various operators
seem to be the ratios $WW/ZZ$ and $\gamma\gamma/\gamma Z$.
The first one, which is applicable in the intermediate
and high Higgs mass range, allows to disentangle $\O_{UB}$
from $\O_{UW}$ down to values of the order of $10^{-1}$, whereas the
second one, applicable in the light Higgs case, is sensitive to
couplings down to the $10^{-3}$ level or less.\par

The identification of CP-violating terms is not directly
possible except for the remark that, contrarily to the
CP-conserving terms, there can be no intereference with the tree
level SM contributions. Thus a CP violating
interaction cannot lower the value of the widths through a
destructive intereference. A direct identification of CP
violation requires
either an analysis of the W or Z spin density matrix through
their fermionic decay distributions \cite{GRS, hzVlachos}, or
the observation of a suitable asymmetry with linearly polarized
photon beams \cite{Kraemer}.

\newpage

\centerline { {\bf Figure Captions }}\par
 Fig.1 Cross sections for Higgs production in
$\gamma\gamma$ collisions
from laser backscattering at a 0.5 TeV (a), 1 TeV (b), 2 TeV (c)
  $e^+e^-$ linear collider. Standard prediction
(solid line),
 with $d =+ 0.01$ (long dashed),  $d =- 0.01$ (dashed - circles),
  $d =+ 0.005$ (short dashed),  $d =- 0.005$ (dashed),
  $d =+ 0.001$ (dashed - stars), and $d =- 0.001$ (dashed - boxes).
 The expected number of events per year
for an integrated luminosity of $20 fb^{-1}$, $80 fb^{-1}$,
$320 fb^{-1}$ is also
indicated. \\
\null\\
 Fig.2 Cross sections for Higgs production in
$\gamma\gamma$ collisions
from laser backscattering at a 0.5 TeV (a), 1 TeV (b), 2 TeV (c)
  $e^+e^-$ linear collider. Standard prediction
(solid line),
 with $\bar d = 0.01$ (long dashed),
  $\bar d = 0.001$ (short dashed),
  $\bar d = 0.005$ (dashed).
 The expected number of events per year
for an integrated luminosity of $20 fb^{-1}$, $80 fb^{-1}$,
$320 fb^{-1}$ is also
indicated. \\
\null\\
 Fig.3  Ratios of Higgs decay widths versus $m_H$,
$\Gamma(H\to\gamma \gamma)/\Gamma(H\to b  \bar b)$ (a),
$\Gamma(H\to\gamma Z)/\Gamma(H\to b  \bar b)$ (b),
$\Gamma(H\to\gamma \gamma)/\Gamma(H\to \gamma Z)$ (c),
$\Gamma(H\to\gamma \gamma)/\Gamma(H\to WW)$ (d),
$\Gamma(H\to\gamma \gamma)/\Gamma(H\to ZZ)$ (e).
 Standard prediction (solid line),
with $d =+ 0.01$ (short dashed), $d =- 0.01$
(dashed),
 $\bar d = 0.01$ (long dashed). \\
\null\\

 Fig.4  Ratios of Higgs decay widths for $m_H=0.2 TeV$
 versus coupling constant values,
$\Gamma(H\to WW)/\Gamma(H\to ZZ)$ (a),
$\Gamma(H\to\gamma \gamma)/\Gamma(H\to \gamma Z)$ (b).
with $d>0 $ (solid), $d<0 $ (short dashed), $d_B>0$
(dashed), $d_B<0$ (long dashed),
 $\bar d $ (dashed-circles), $\bar d_B$ (dashed-stars).

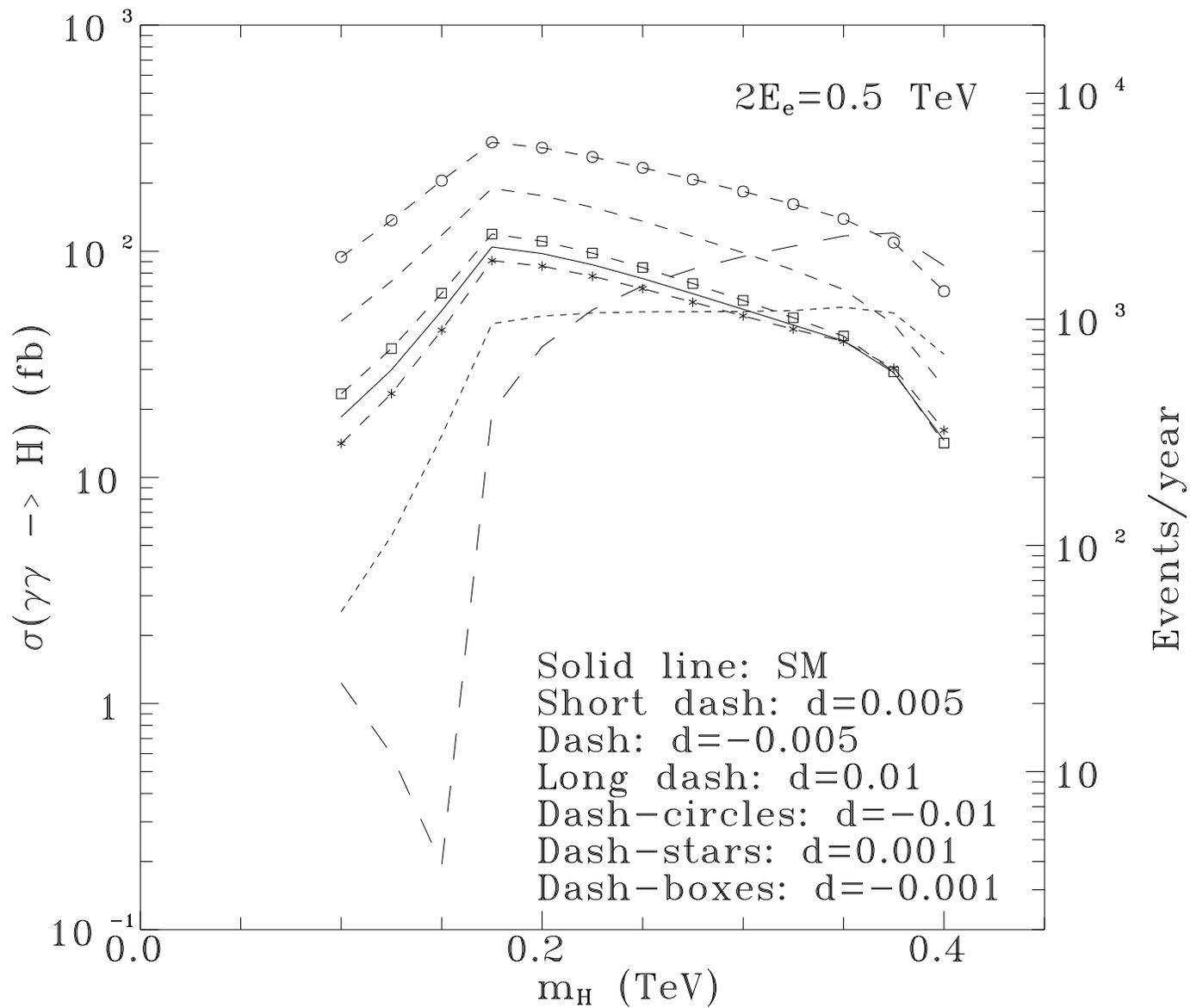
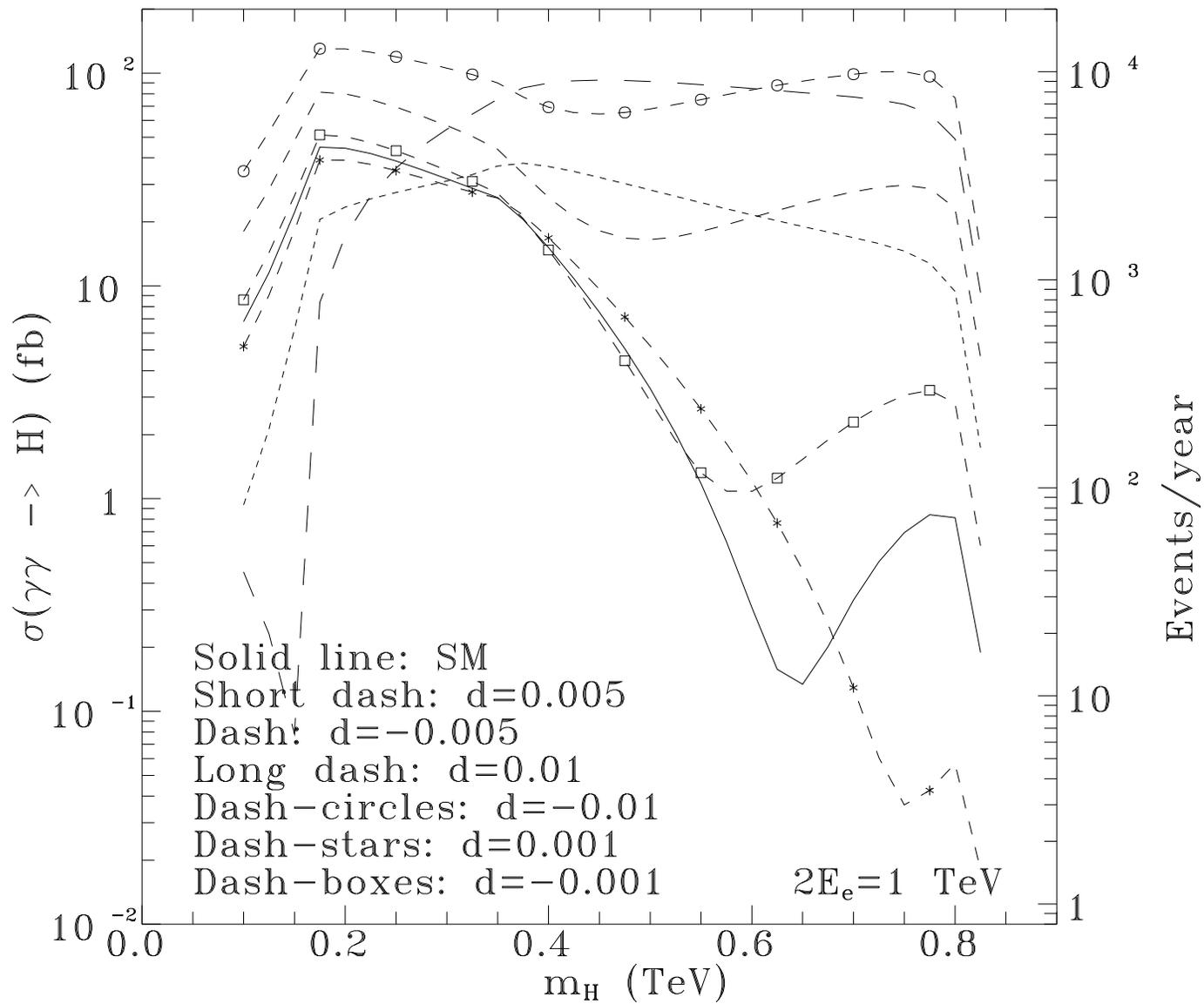
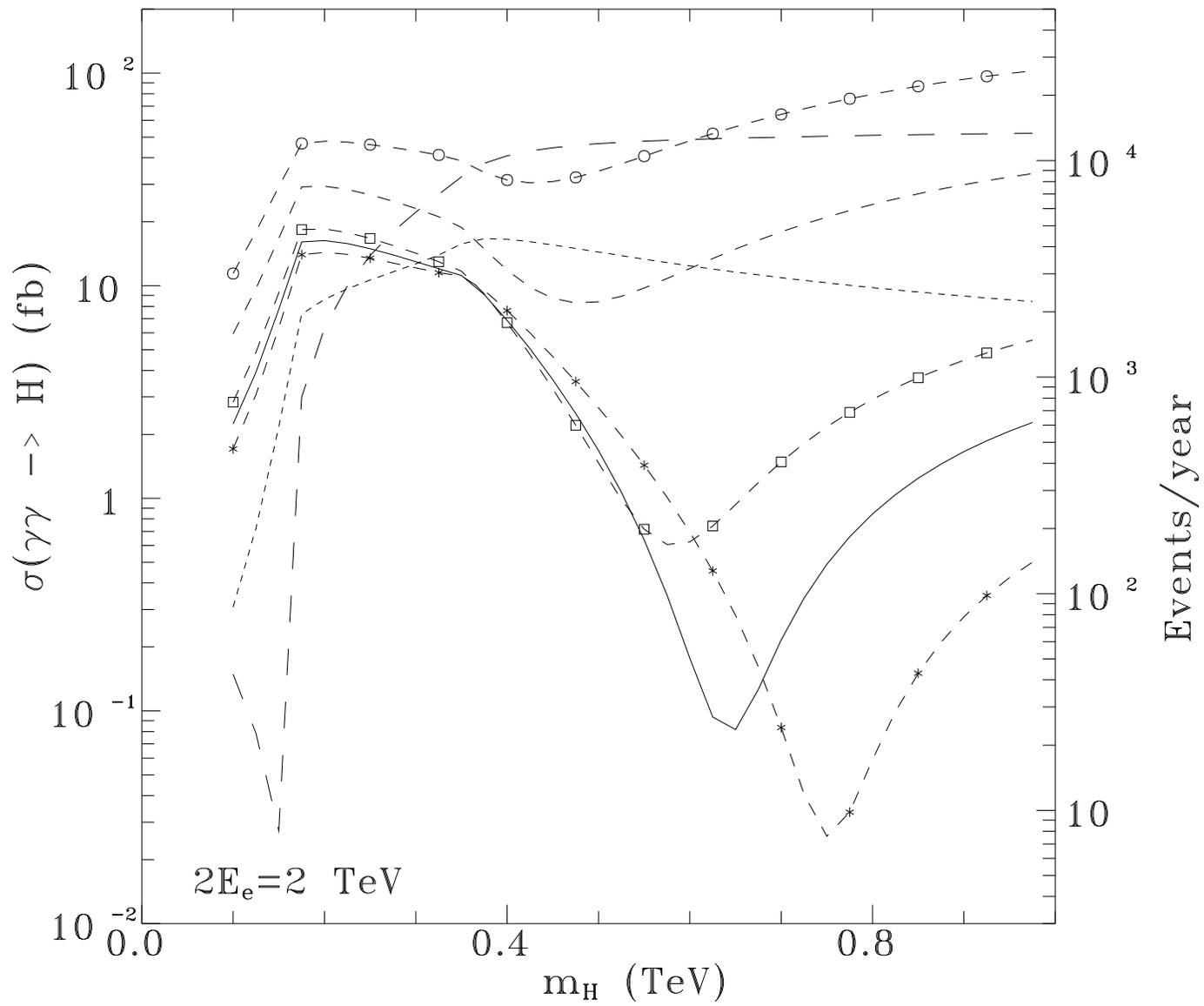
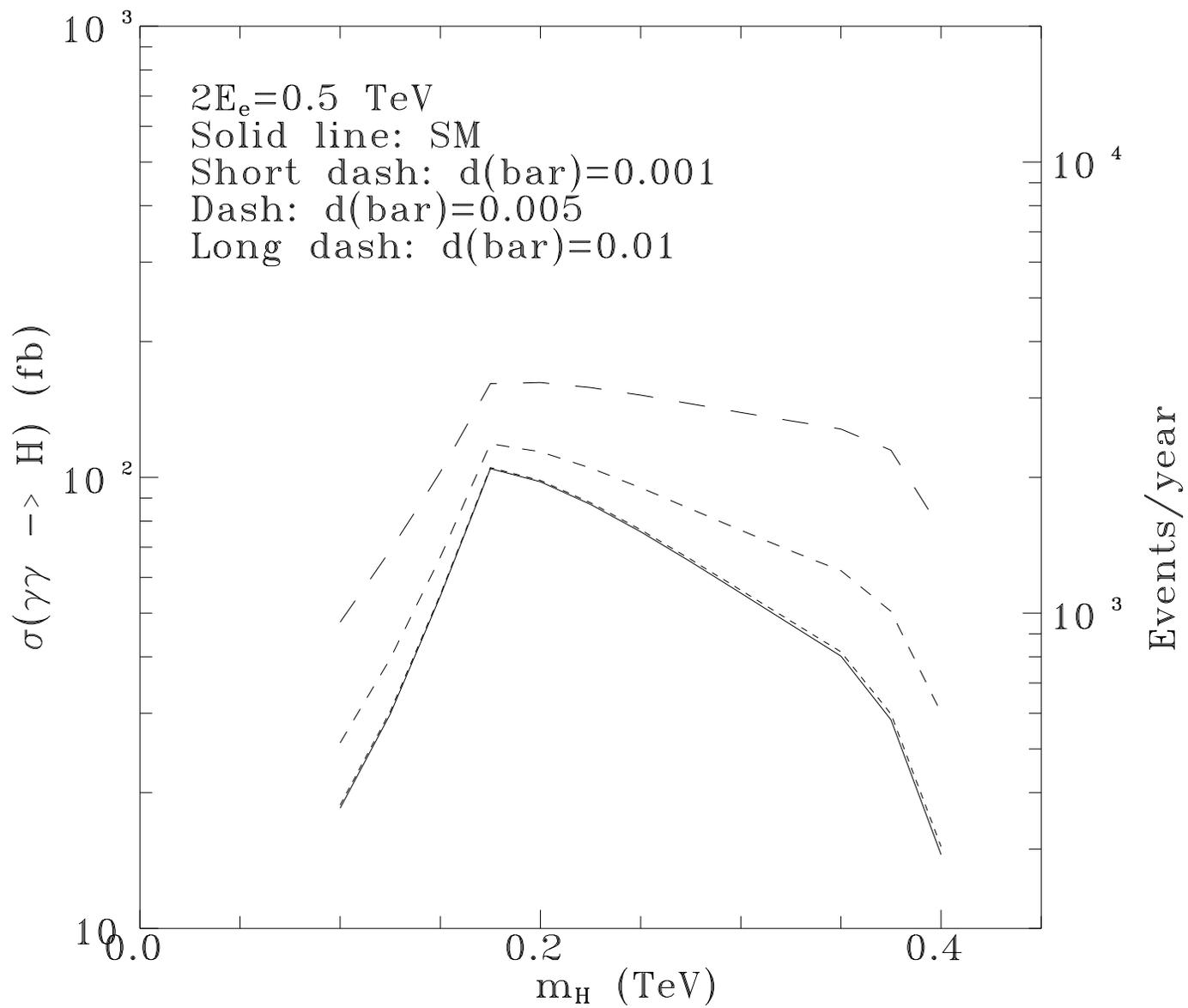
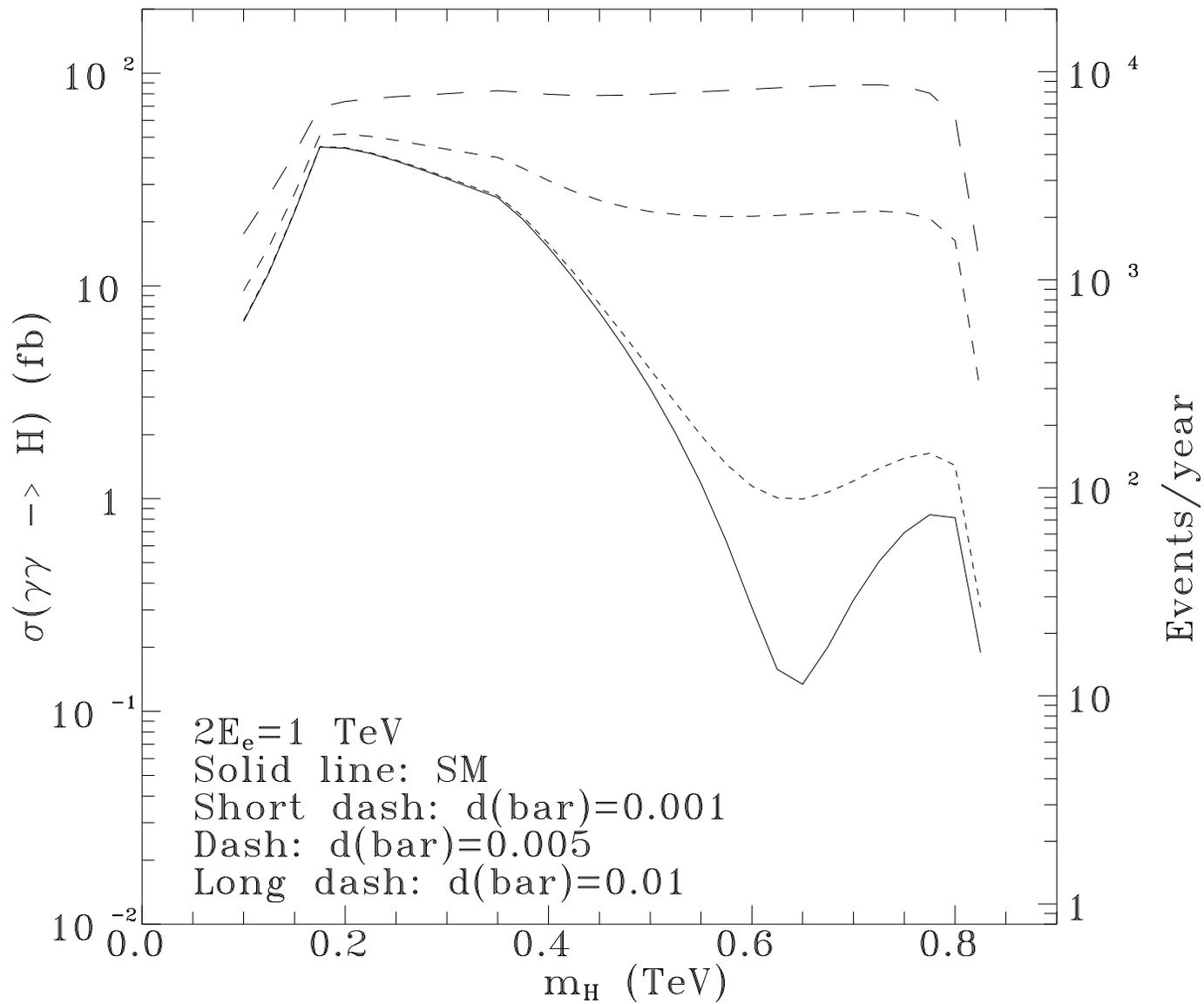
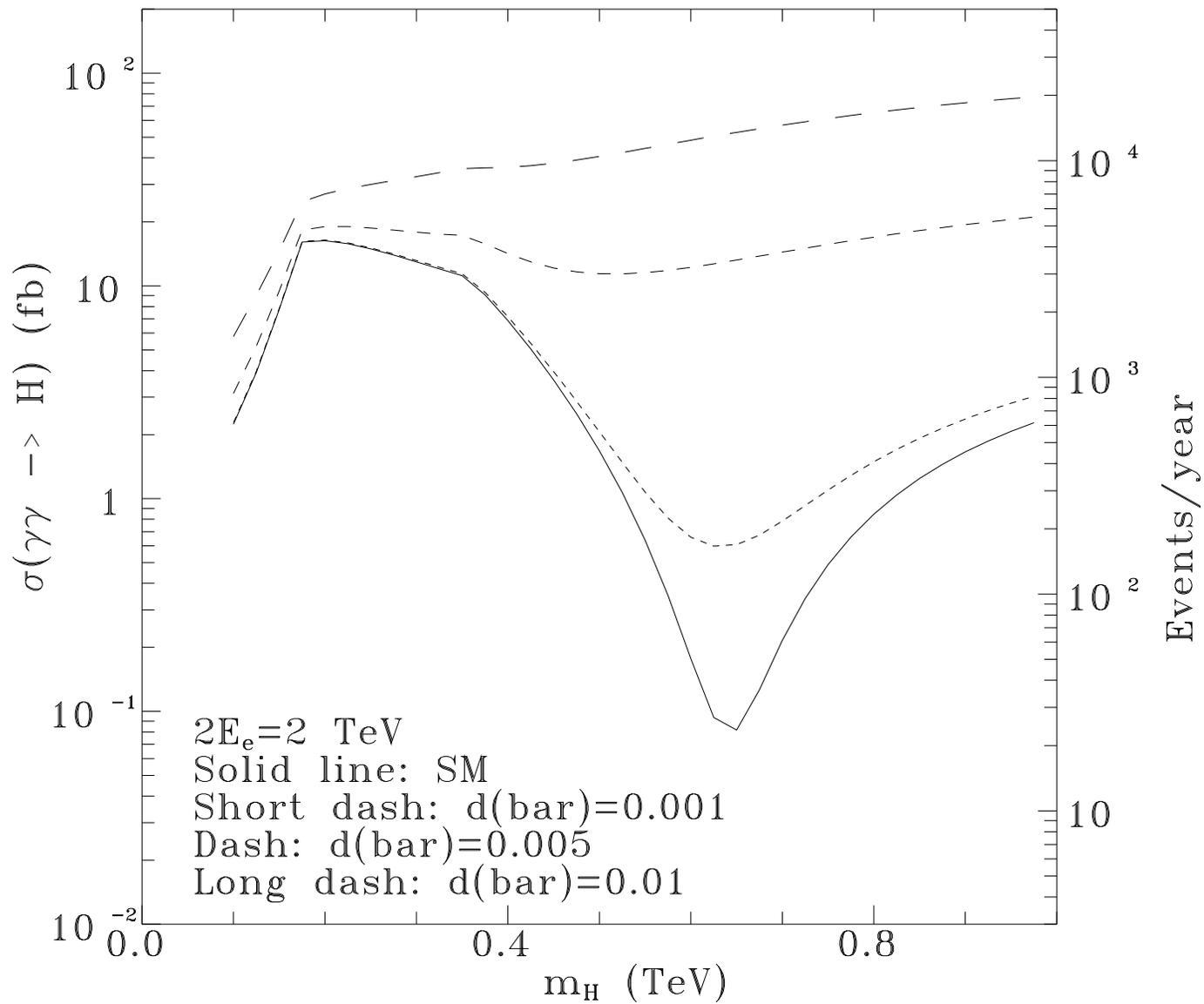
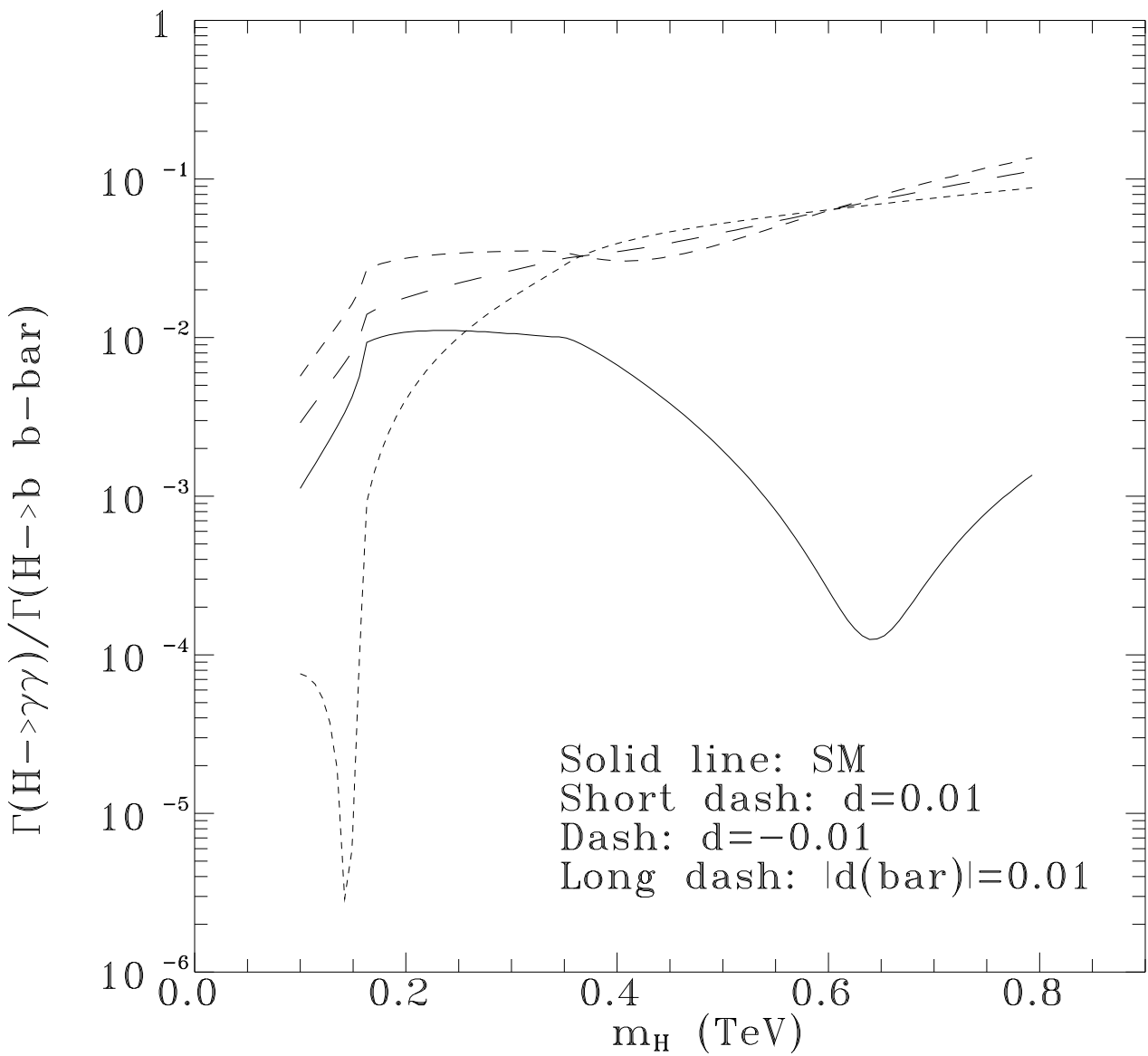
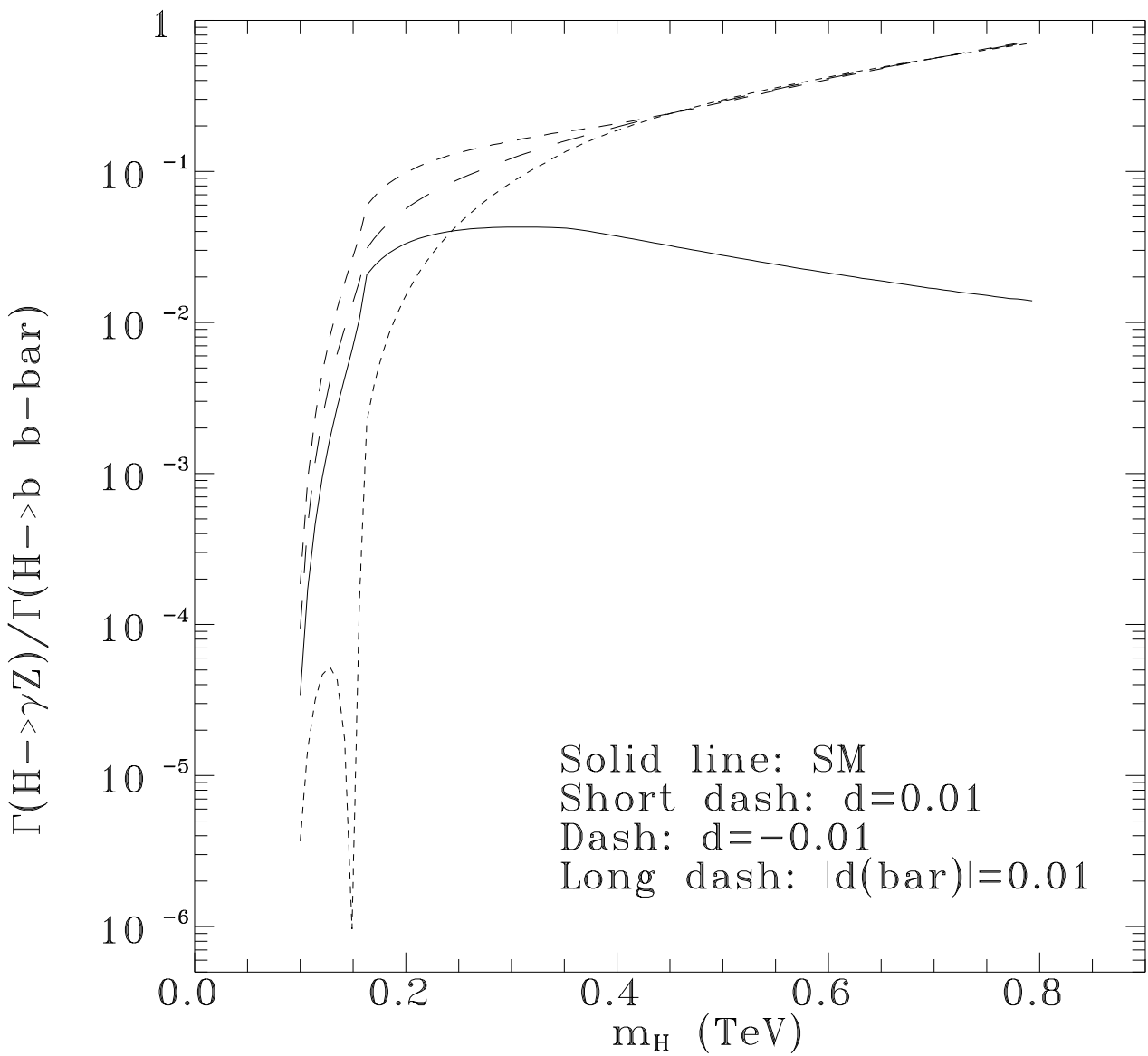
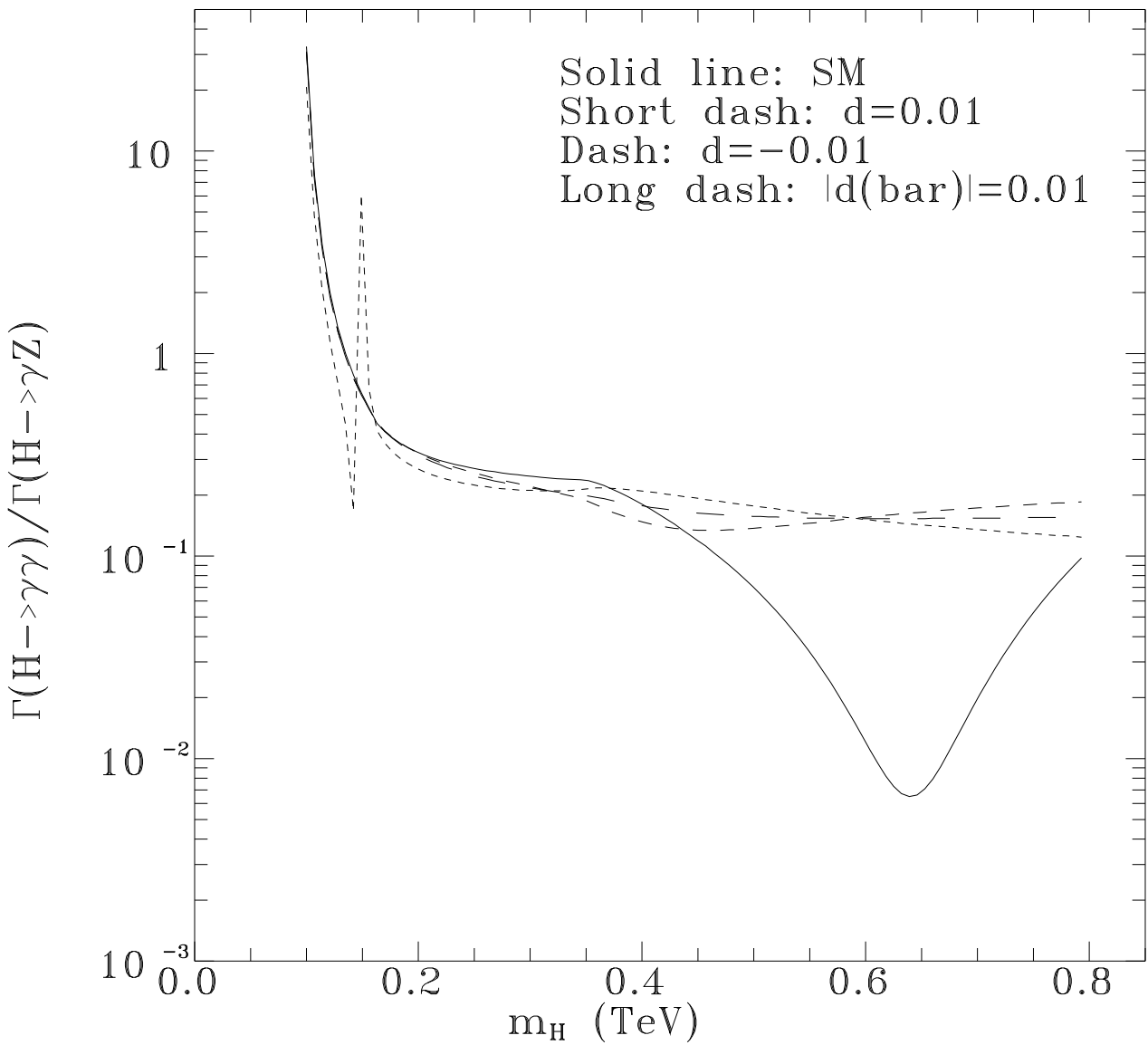
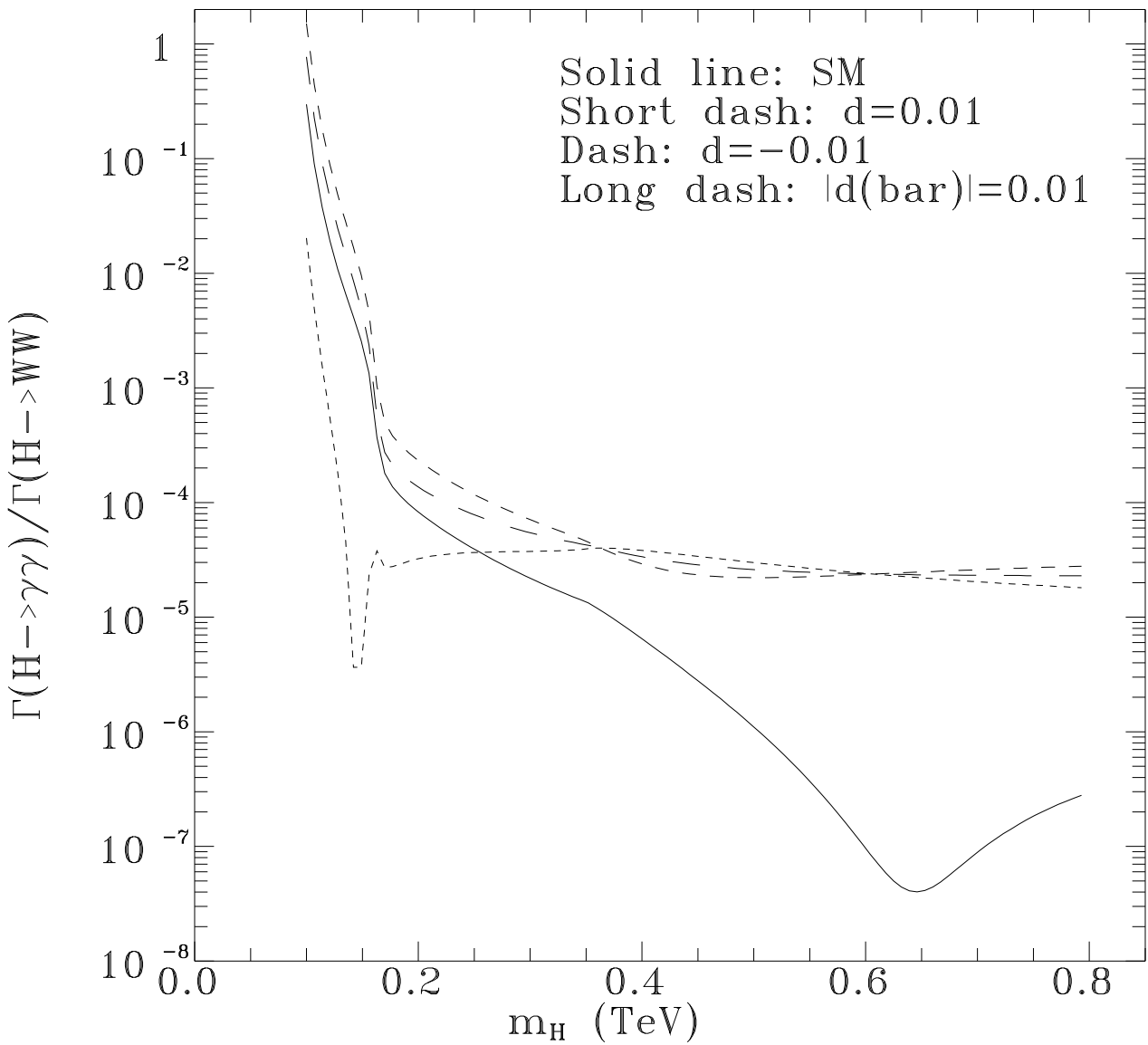
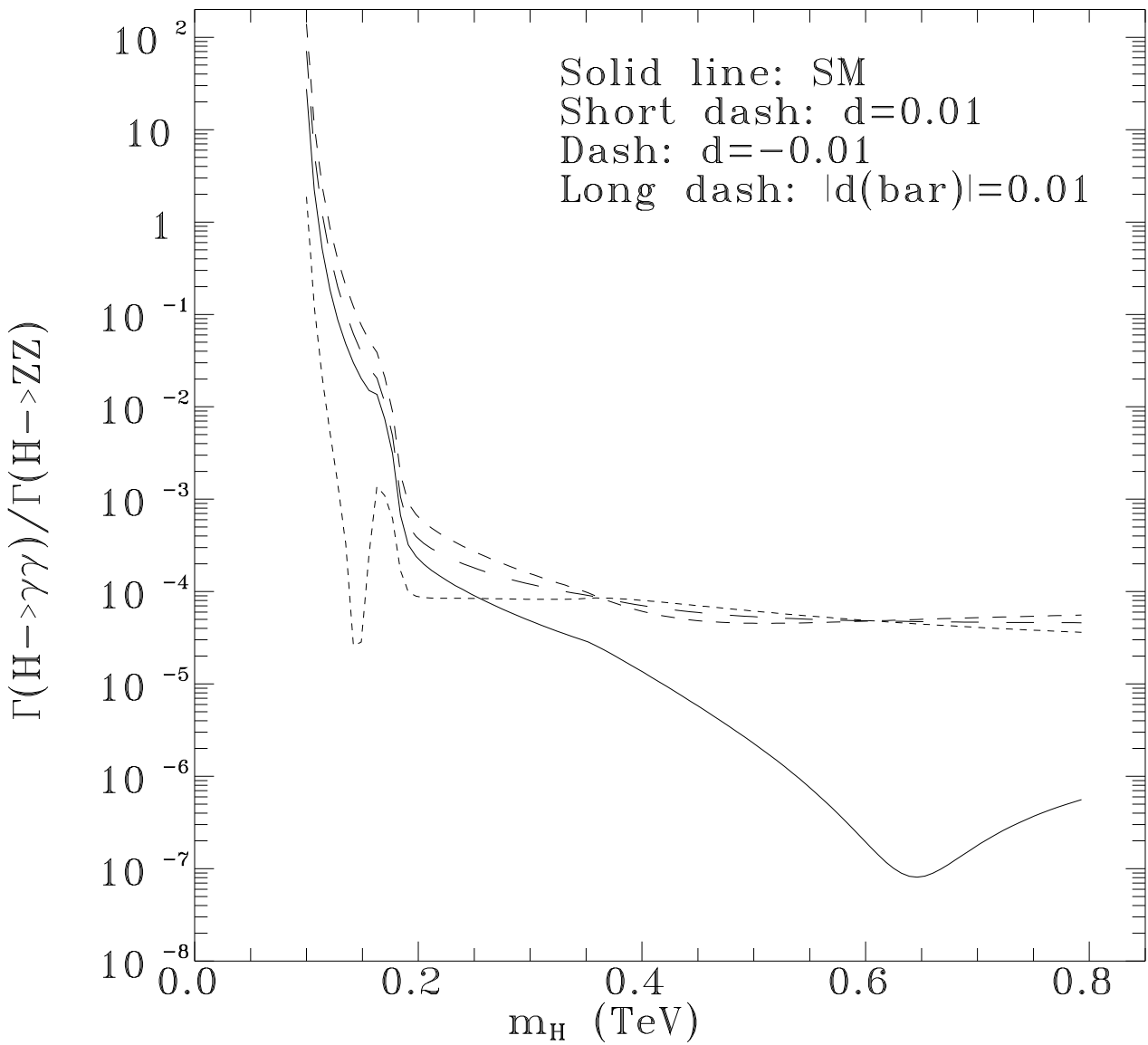
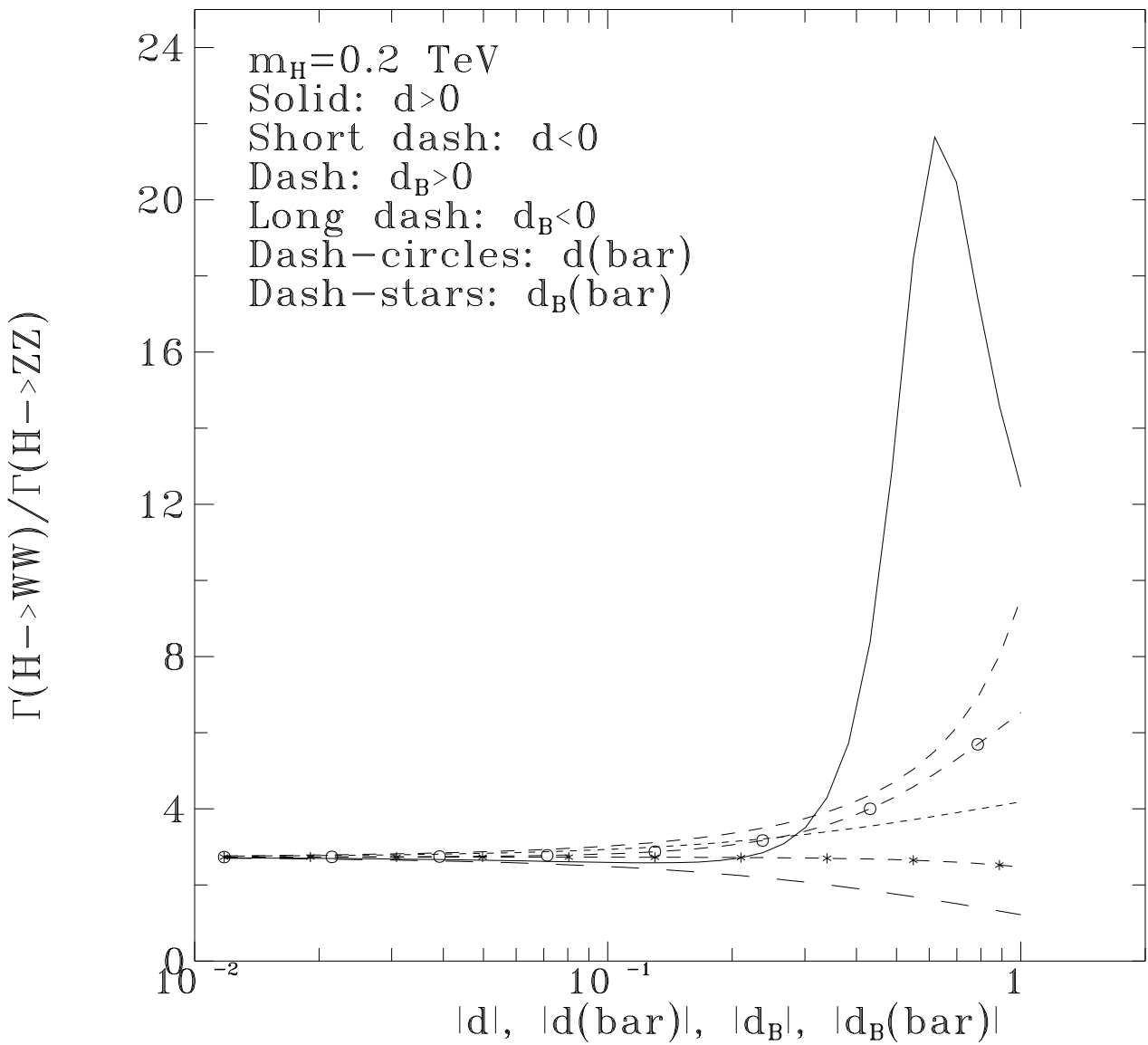
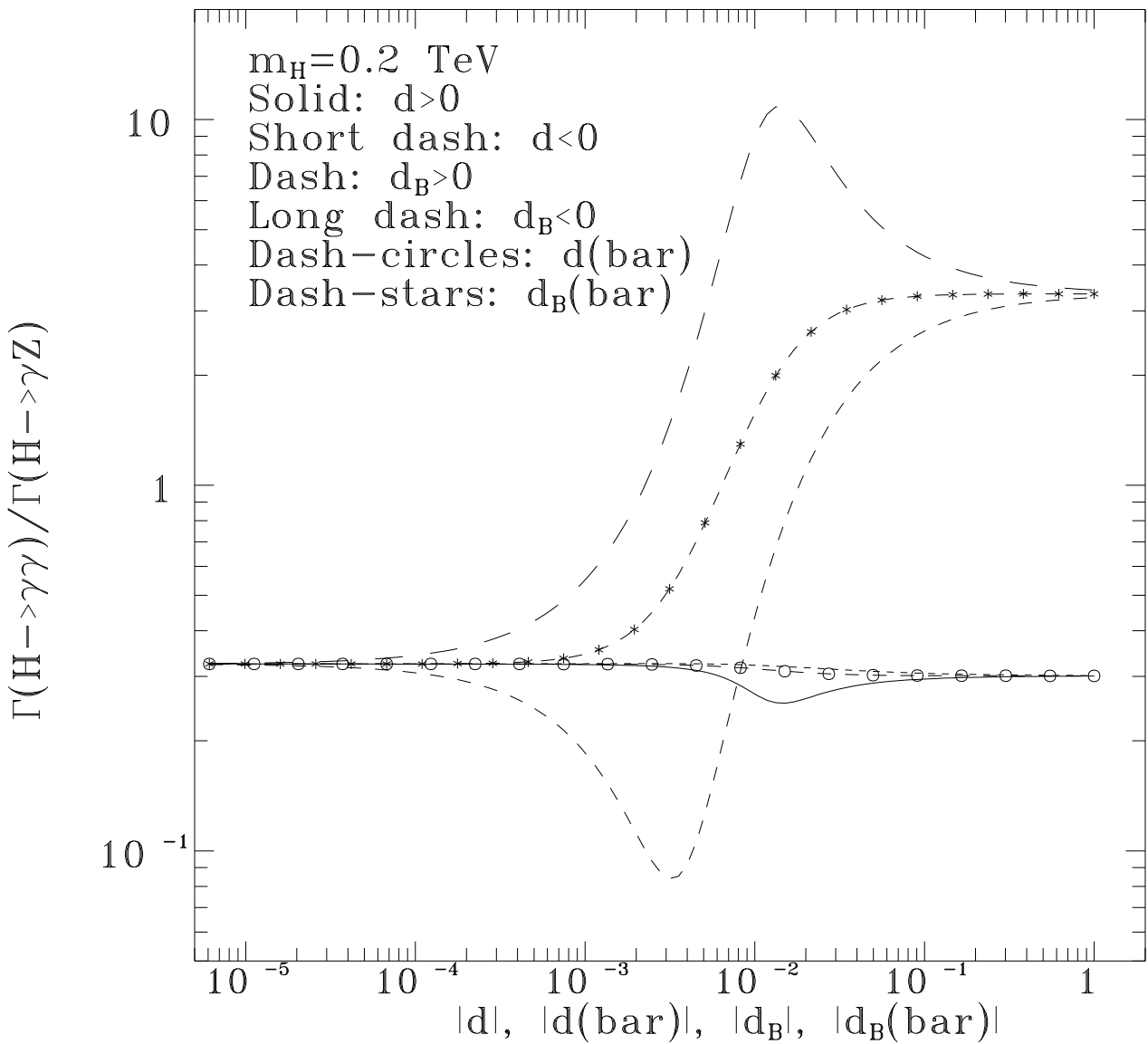
\end{document}